\documentclass[12pt,preprint]{aastex}
\shorttitle{Bullet Cluster in cDE Models}
\shortauthors{Lee \& Baldi}
\begin{document}
\title{MASSIVE GRAVITY WRAPPED IN THE COSMIC WEB}
\author{Junsup Shim\altaffilmark{1}, Jounghun Lee\altaffilmark{1}, 
Baojiu~Li\altaffilmark{2}}
\altaffiltext{1}{Astronomy Program, Department of Physics and Astronomy, FPRD, 
Seoul National University, Seoul 151-747, Korea\\ 
\email{jsshim@astro.snu.ac.kr, jounghun@astro.snu.ac.kr}}
\altaffiltext{2}{Institute of Computational Cosmology, Department of Physics, 
Durham University, Durham DH1 3LE, UK}
\begin{abstract}
We study how the filamentary pattern of the cosmic web changes if the true gravity deviates from the general 
relativity (GR) on the large scale. The $f(R)$ gravity whose strength is controlled to satisfy the current 
observational constraints on the cluster scale is adopted as our fiducial model and a large $N$-body simulation of 
high-resolution is utilized for this study. By applying the minimal spanning tree algorithm to the halo catalogs from 
the simulation at various epochs, we identify the main stems of the rich superclusters located in the most 
prominent filamentary section of the cosmic web  and determine their spatial extents per member cluster as the 
degree of their straightness.  It is found that the $f(R)$ gravity has an effect of significantly bending the 
superclusters and that the effect becomes stronger as the universe evolves. Even in the case where the deviation 
from GR is too small to be detectable by any other observables, the degree of the supercluster straightness 
exhibits conspicuous difference between the $f(R)$ and the GR models.  Our results also imply that the 
supercluster straightness could be a useful discriminator of $f(R)$ gravity from the coupled dark energy since it is 
shown to evolve differently between the two models. As a final conclusion, the degree of the straightness of the 
rich superclusters should provide a powerful cosmological test of large scale gravity.
\end{abstract}
\section{INTRODUCTION}

The deepest and the most profound question in modern cosmology is what caused the universe to 
accelerate at the present epoch. Although the Planck mission team recently confirmed the stunning 
agreements between the predictions of the standard $\Lambda$CDM (cosmological constant $\Lambda$
+cold dark matter) model and the Cosmic Microwave Background (CMB) temperature power spectrum 
measured with unprecedentedly high precision \citep{planck-xv13,planck-xvi13}, the notorious fine tuning 
problem of $\Lambda$ still haunts the cosmologists to vigorously look for alternative models.  There have been 
two main directions in developing viable alternatives. One direction is  to replace $\Lambda$ with some dynamic 
dark energy with negative pressure that could induce the current acceleration of the universe without requiring fine 
tuned conditions \citep[see][for a comprehensive review]{AT10}. Among various dynamic dark energy scenarios,  
the coupled dark energy (cDE) model where a scalar field dark energy interacts with dark matter has been found 
quite promising because of its capacity of alleviating several reported tensions between the $\Lambda$CDM model 
and the observations \citep[e.g.,][]{BLM11,baldi12a,LB12,SM13}.

The other main direction is to modify the general relativity (GR) on the large scale, which makes the concept 
of anti-gravitational dark energy unnecessary to explain the observed distance-luminosity relation of Type Ia 
supernovae \citep[see][for a comprehensive review]{clifton-etal12}. The tremendous success of GR on the local scale, 
however,  leaves only very little room for possible deviation of true gravity from GR.  
The $f(R)$ gravity \citep[e.g.,][and references therein]{HS07,LB07} is one of those few modified gravity models which 
has so far survived severe cosmological tests \citep{reyes-etal10,wojtak-etal11}.  
In this model,  $f(R)$ represents an arbitrary function of the Ricci scalar $R$ that is substituted for $R$ in the Einstein-
Hilbert action, and its derivative called the scalaron, $df/dR$,  induces a fifth force on the large scale, the strength of 
which is quantified by its absolute magnitude  at the present epoch, $f_{R0}\equiv |df/dR|_{0}$ \citep{SF10,DT10}.  
An essential feature of the $f(R)$ gravity is the presence of the chameleon mechanism that blocks deviation of gravity 
from GR in dense environment: The denser the environment is, the weaker the fifth force is \citep[e.g.,][]{KW04,MS07}. 

Although the abundance of galaxy clusters and the strength of their gravitational clusterings have been widely 
used as one of the most powerful probes of the background cosmology \citep[for a review, see][]{allen-etal11}, 
these probes are unlikely to be efficient discriminators of modified gravity, since the galaxy clusters are usually 
located in the highly dense supercluster environments where the chameleon effect should be very strong.  
A recent trend in the cosmological study of $f(R)$ gravity is to explore its effect on the low-density regions and to figure 
out which observables among the low-density phenomena is the best indicator of $f(R)$ gravity. For instance, the 
dynamic mass of field galaxies, the spin parameters of dwarf void galaxies, the abundance of cosmic voids, 
and etc. have been suggested as useful indicators of large-scale gravity 
\citep{zhao-etal11a,lee-etal13,clampitt-etal13}. 

Very recently, \citet{SL13} showed by analyzing the halo catalogs from N-body simulations that the degree of the 
straightness of rich superclusters changes significantly by the presence of cDE. Their results are summarized as 
(i)  the superclusters tend to be less straight in cDE models with stronger coupling; (ii) the difference in the degree 
of the supercluster straightness is much larger than that in the abundance of the clusters (or superclusters) among 
different cDE models; (iii) the difference is larger at higher redshifts.  \citet{SL13} provided the following 
explanations for their result: 
The fifth force generated by the coupling between dark matter and dark energy in cDE models plays a role in making 
the gravitational clustering of galaxy clusters less isotropic, which is best manifested by the straightness of the 
superclusters that  correspond to the most prominent filamentary structures of the cosmic web. 

It is intriguing to ask if the long-range fifth force generated by the scalaron in the $f(R)$ gravity model also affects on 
the supercluster straightness. In fact, given the result of \citet{SL13}, it is reasonable to expect that the superclusters 
should be less straight in $f(R)$ gravity models than in the $GR$ model. The essential work to undertake here is to 
investigate quantitatively how sensitively the degree of the supercluster straightness changes by the presence of $f(R)$ 
gravity and to examine whether or not it would be powerful enough to distinguish $f(R)$ not only from GR but 
also from cDE. 

The contents of the upcoming sections are outlined as follows. In section \ref{sec:data} are briefly described the data 
from N-body simulations for $f(R)$ gravity and the algorithms employed to determine the superclusters and their 
degree of straightness. In section \ref{sec:results} are presented the mains result on the dependence of the degree of 
the supercluster straightness on the strength of $f(R)$ gravity. In section \ref{sec:conclusion} is drawn a final conclusion. 

\section{DATA AND ALGORITHM}\label{sec:data}

To run a N-body simulation for a $f(R)$ gravity model, it is first necessary to specify the function $f(R)$. 
We adopt the following Hu-Sawicki model characterized by two parameters $n$ and $c_{1}/c_{2}$ 
\citep{HS07}: 
\begin{equation}
\label{eqn:fr}
f(R)=-m^{2}\frac{c_{1}(-R/m^{2})^{n}}{c_{2}(-R/m^{2})^{n}+1}, \
\end{equation}
Here, $m\equiv 8\pi G\bar{\rho}_{m}/3$ where $\bar{\rho}_{m}$ represents the mean mass density of the 
universe at present epoch.  Following the previous works \citep{oyaizu08,zhao-etal11b}, the two parameter 
values are set at $c_{1}/c_{2}=6\Omega_{\Lambda}/\Omega_{m}$ and $n=1$. 
The comparison between the observed abundance evolution of galaxy clusters and the analytic mass function
has yielded a tight constraint of $|f_{R0}|\lesssim 10^{-4}$  for the Hu-Sawicki $f(R)$ model 
\citep{schmidt-etal09,lombriser-etal10}. Given this cluster-scale constraint, we consider three models: GR and 
two $f(R)$ gravity models, F5 and F6, for which the values of $|f_{R0}|$ are set at $10^{-5}$ and $10^{-6}$, 
respectively. Throughout this paper, GR represents the standard $\Lambda$CDM cosmology where the gravity is 
described by GR.

For each model, we run a large $N$-body simulation by employing the ECOSMOG code \citep{li-etal12b}. 
The simulation contains a total of $1024^{3}$ dark matter particles in its periodic box of linear size $1\,h^{-1}$Gpc.   
The initial conditions for each model are all tuned by setting the key cosmological parameters at  
$\Omega_{m}=0.24,\ \Omega_{\Lambda}=0.76,\ \Omega_{b}=0.045,\ h=0.73,\ \sigma_{8}=0.8,\ n_{s}=0.96$.
The Amiga's Halo Finder (AHF) code \citep{KK09} are utilized to identify the  bound halos from the spatial 
distributions of the dark matter particles. 
For the detailed description of the simulations and the halo-identification procedures, 
see \citet{KK09} and \citet{li-etal12a,li-etal12b}. 

Two additional algorithms are employed for our analysis: the friends-of-friends (FoF) group finder and the 
minimal spanning tree (MST) algorithm. The former is used for the identification of the superclusters and 
the latter detects the interconnection among the member clusters of each supercluster. Both of the codes 
treat the cluster-size halos as particles without weighting them by their mass. In other words, no information 
on the masses of cluster-size halos is required to apply the two algorithms. Only the lower cut-off mass has 
to be specified when the cluster-size halos are selected (see section \ref{sec:results}). 
It is worth mentioning here that the AHF  which is used to identify the bound halos is not appropriate 
to find the superclusters since the AHF is basically a refined spherical over density algorithm \citep{KK09} while 
the superclusters are well known to have filamentary shapes 
\citep[e.g.,][]{dekel-etal84,west89,plionis-etal92,jaaniste-etal98,basilakos-etal01,basilakos03,einasto-etal07,
wray-etal06,einasto-etal11}. 

The MST technique has been widely used to understand the interconnected structures of the cosmic web 
\citep[e.g.,][]{barrow-etal85,KS96,doroshkevich-eta01,colberg07,PL09a,PL09b,SL13}. 
As mentioned in \citet{SL13}, its usefulness lies in the fact that it does not require to know the underlying distribution of 
dark matter particles and thus can be directly applied to the observed spatial distributions of galaxies or clusters. 
In the following section, we describe in detail how the degree of the supercluster straightness is measured from 
the data with the help of the above algorithms and how it is different among the three models, GR, F6 and F5. 
Five different epochs will be considered: $a=1.0,\ 0.9,\ 0.8,\ 0.7,\ 0.6$ where $a$ is the scale factor.

\section{EFFECT OF f(R) GRAVITY ON THE SUPERCLUSTER STRAIGHTNESS}
\label{sec:results}

We take the same procedures that \citet{SL13} followed to determine the degree of the supercluster 
straightness at each epoch for each model:  
\begin{itemize}
\item
Select those halos with mass $M_{\rm c}\ge 10^{13}\,h^{-1}M_{\odot}$ as the clusters and identify the 
FoF groups of clusters as the superclusters. As conventionally done when the marginally bound superclusters 
are identified as the FoF groups \citep[e.g.,][]{wray-etal06,KE05,LP06,LE07,SL13}, the linking length is set at 
one third of the the mean separation distance among the selected clusters.
\item
Find the MSTs of those rich superclusters with $N_{\rm c}\ge 3$ where $N_{\rm c}$ is the number of the member 
clusters (nodes) and prune each supercluster MST to determine its main stem, called a "spine" by \citet{SL13}. 
\item
Select only those rich superclusters with $N_{\rm node}\ge 3$ after the pruning where $N_{\rm node}$ denotes the 
number of the nodes that make up the spine of each supercluster.  
\item 
Measure the size of each supercluster spine as 
$S=\left[\sum_{i=1}^{3}(x_{i,{\rm max}}-x_{i,{\rm min}})^{2}\right]^{1/2}$ 
when the comoving Cartesian coordinates of a node, $\{x_{i}\}_{i=1}^{3}$, is in the range of 
$x_{i, {\rm min}}\le x_{i}\le x_{i, {\rm max}}$. 
\item
Determine the degree of the straightness of each supercluster spine as its specific size defined as, 
$\tilde{S}=S/N_{\rm node}$  by \citet{SL13} and then take the average of $\tilde{S}$ over all the selected supercluster 
spines. 
\end{itemize}

\begin{figure}[ht]
\begin{center}
\plotone{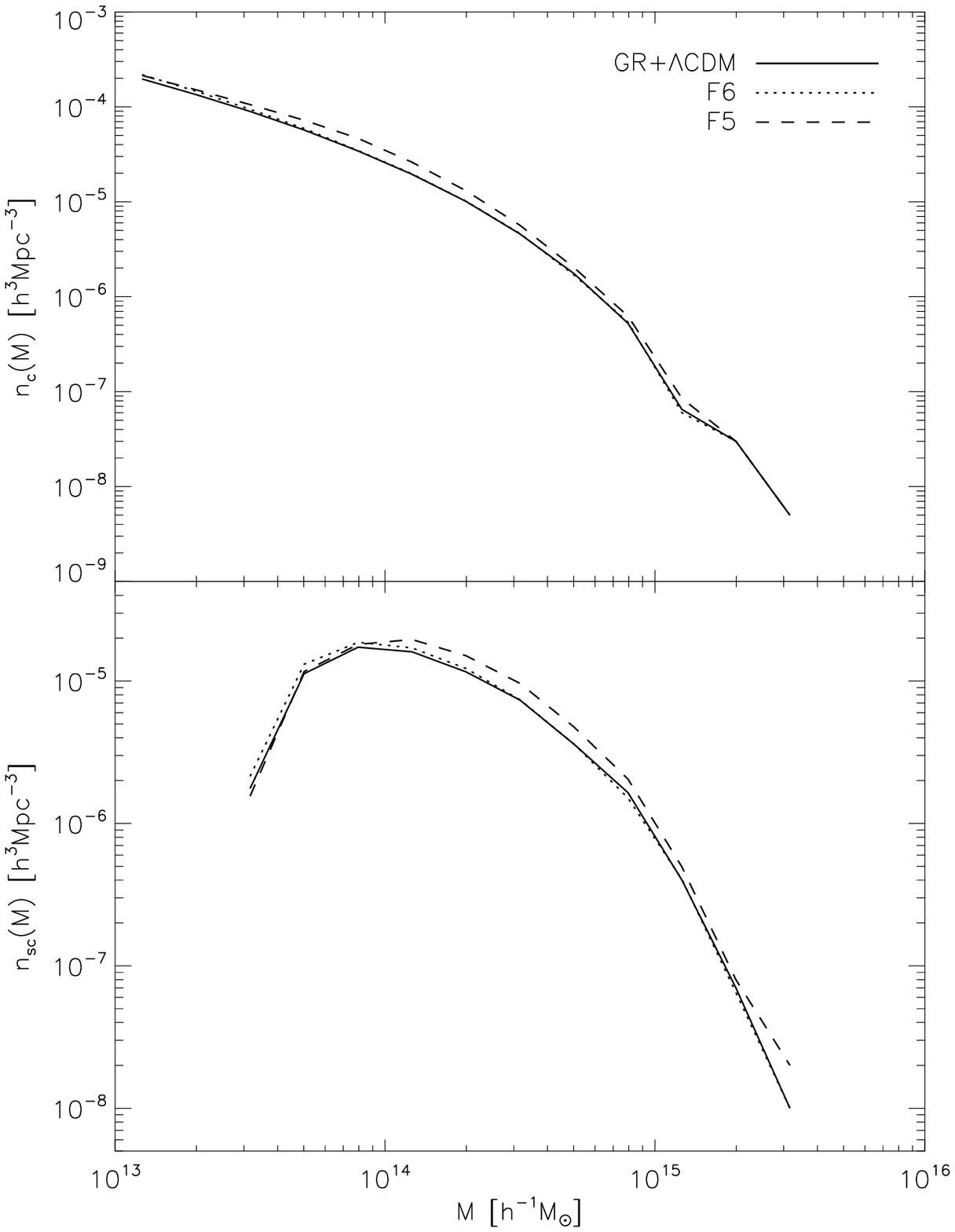}
\caption{Mass functions of the cluster and the supercluster halos at $z=0$ for three different gravity models 
in the top and bottom panels, respectively.}
\label{fig:mf}
\end{center}
\end{figure}

Figure \ref{fig:mf} plots the mass functions of the clusters (top panel) and the superclusters (bottom panel) 
at the present epoch ($a=1.0$)  for three different models. As can be seen, the mass functions have highest amplitudes 
in the F5 model while there is almost no difference between the GR and the F6 cases.  Table \ref{tab:spine} lists the 
number of the supercluster spines which consist of three or more nodes and the average values of their specific masses 
defined as $M_{\rm spine}/N_{\rm node}$  where $M_{\rm spine}$ is the sum of the masses of all the nodes of a 
supercluster spine. 

\begin{figure}[ht]
\begin{center}
\plotone{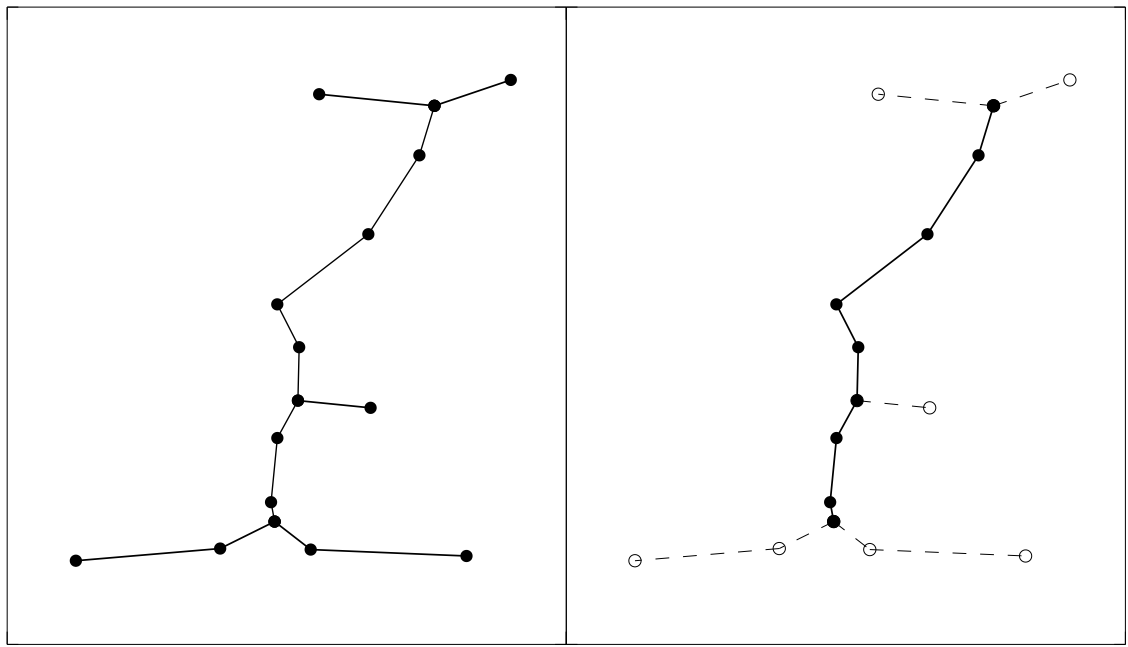}
\caption{Pruning of a supercluster to determine its main stems (spine).}
\label{fig:prune}
\end{center}
\end{figure}

\begin{figure}[ht]
\begin{center}
\plotone{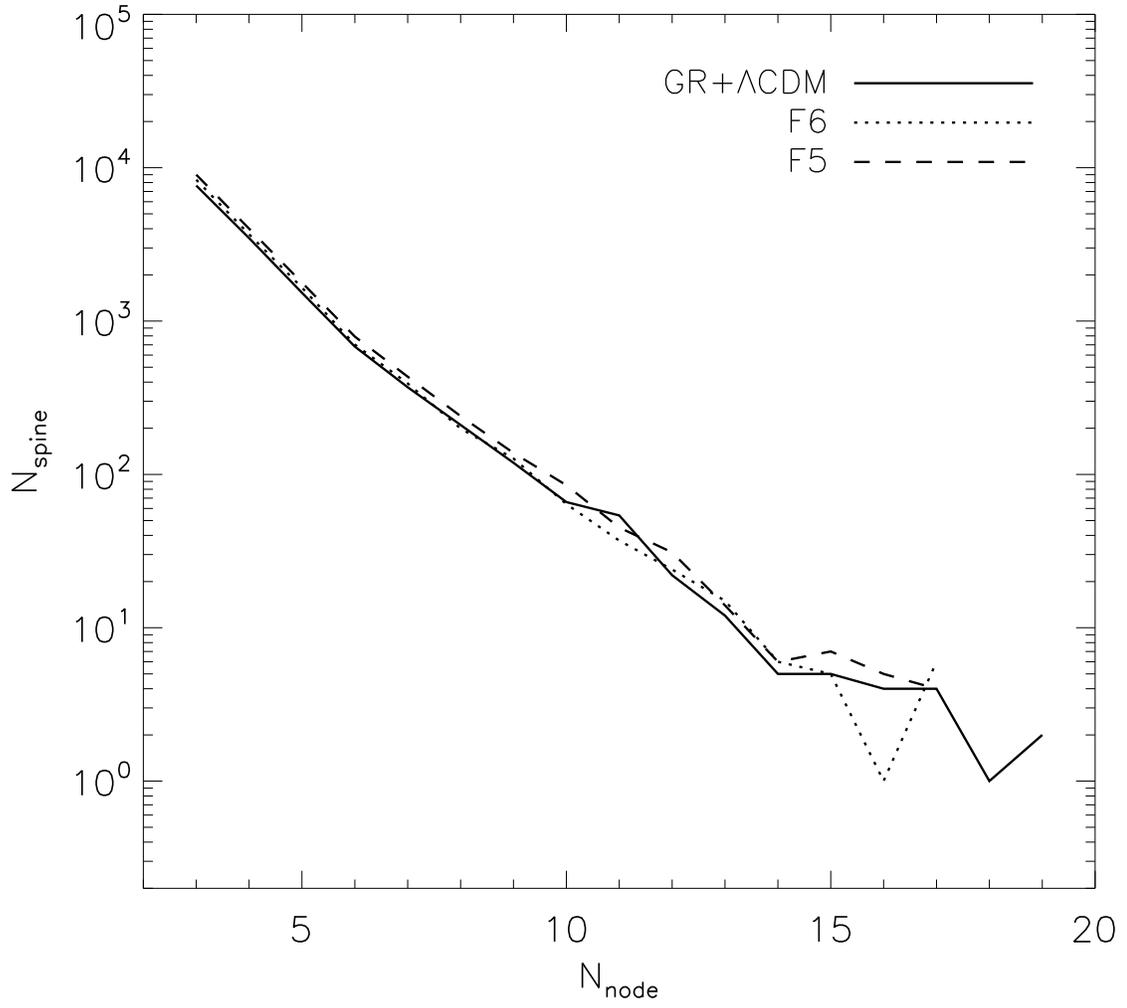}
\caption{Number distribution of the superclusters as a function of node at $z=0$ for three different models.}
\label{fig:nf}
\end{center}
\end{figure}

In Figure \ref{fig:prune}, the pruning process of a supercluster MST  is depicted at $z=0$ for the GR case 
in the two dimensional plane projected onto the $x_{1}$-$x_{2}$ plane. In the left panel the solid line 
represents a supercluster MST before pruning and the dots connected by the solid line correspond to their nodes. 
In the right panel, the solid line corresponds to a supercluster spine (i.e., MST after pruning) while the dashed lines 
represents the minor branches pruned away from the MST. For the detailed explanation about the pruning process, 
please see \citet{colberg07}. 
Figure \ref{fig:nf} plots  the number distributions of the supercluster spines, $N_{\rm spine}$ vs. 
the number of nodes, $N_{\rm node}$ at the present epoch for three models. The result shows 
that other than the numerical fluctuations there is almost no difference in the node number distribution of the 
supercluster spines between the GR and the F6 models, while the F5 model has a noticeably higher amplitude. 

\begin{figure}[ht]
\begin{center}
\plotone{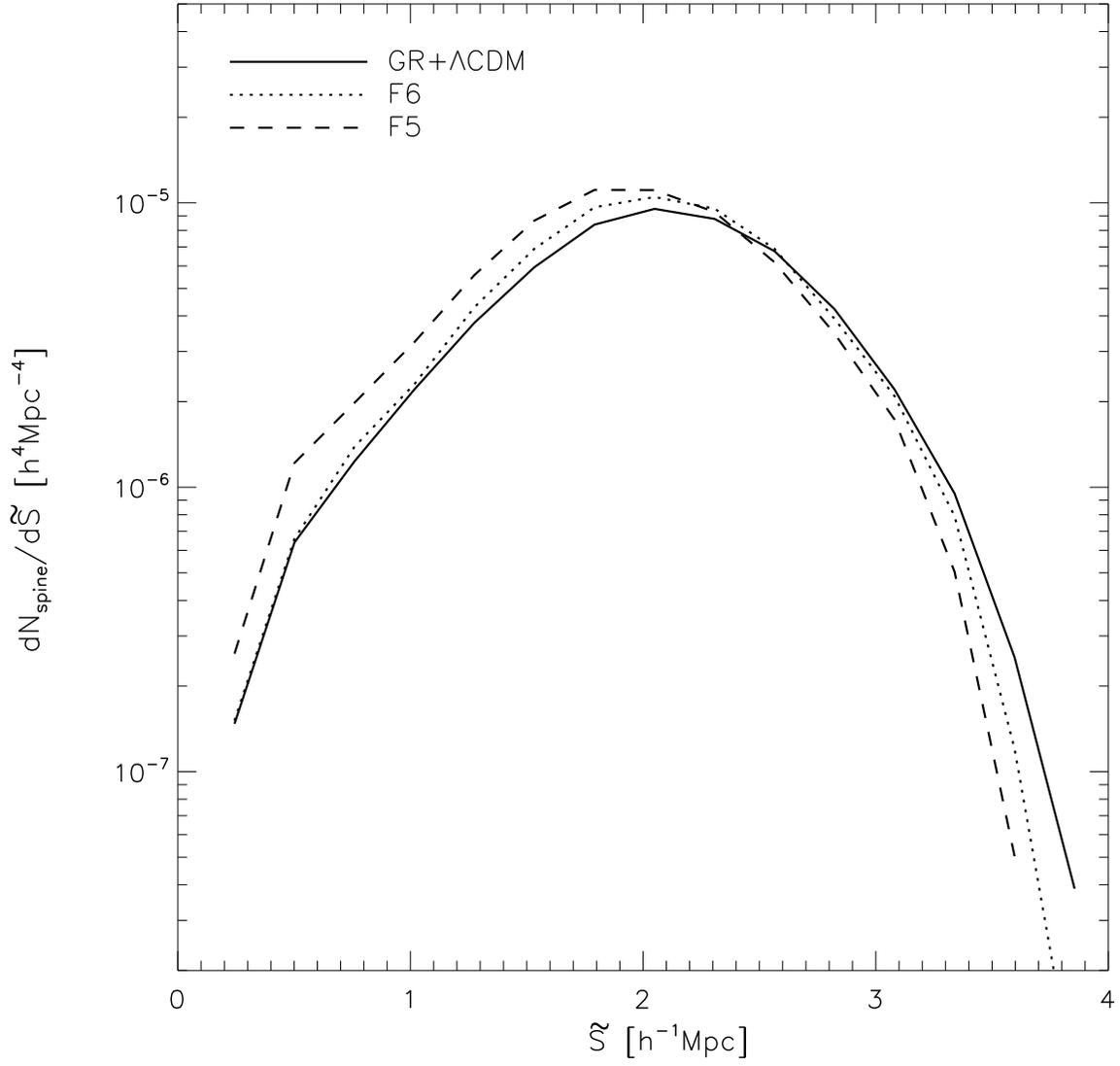}
\caption{Specific size distributions of the superclusters at $z=0$ for three different models.}
\label{fig:sf}
\end{center}
\end{figure}

Figure \ref{fig:sf} plots the specific size function defined as $dN_{\rm spine}/d\tilde{S}$ per unit volume 
at the present epoch for the three models. As can be seen, there is a noticeable difference in the specific size function 
among the three models. The specific size function has the highest (lowest) amplitude in the GR (F5) case.  
Note that there is appreciable difference in the specific size function even between GR and F6. 
The comparison with the results shown in Figure \ref{fig:mf} reveals that the difference in the specific size functions 
among the three models is much bigger than that in the mass function of the superclusters. In other words, the 
specific size function of the supercluster spine should be much better indicator of large-scale gravity. 

\begin{figure}[ht]
\begin{center}
\plotone{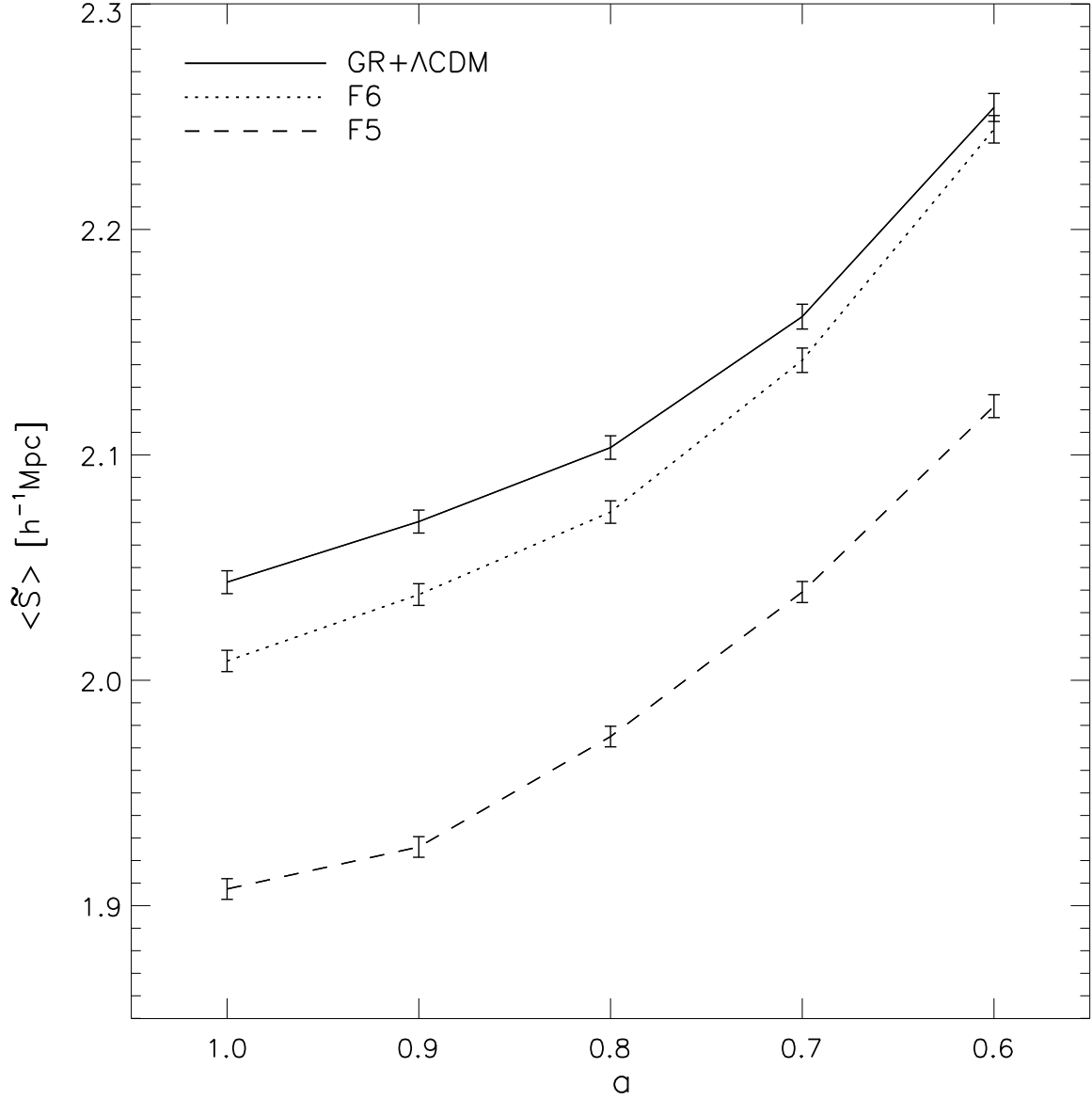}
\caption{Evolution of the mean specific sizes of the superclusters with the scale factor $a$ for three different 
models.}
\label{fig:ms}
\end{center}
\end{figure}

Figure \ref{fig:ms} shows the average specific sizes of the supercluster spines vs. the scale factor for the three 
models, demonstrating how $\langle\tilde{S}\rangle$ evolves in each model. The errors are calculated as the one 
standard deviation in the measurements of the averages. 
As can be seen, the mean specific sizes of the superclusters in the GR (F5) model has the highest (lowest) values at 
all epochs.  Note that there is a significant difference between the F6 and the GR models in the average specific 
size of the supercluster spines although it looks small compared with that between the F5 and the GR models. 
This result implies that the degree of the straightness of the superclusters should be useful as a new cosmological 
test of gravity. 

Note also that the difference in $\langle\tilde{S}\rangle$ between the F6 and the GR model increases as the 
universe evolves. In other words, the effect of $f(R)$ gravity on the degree of the supercluster straightness 
becomes stronger as the Universe evolves. It is interesting to compare this result with that of 
of \citet{SL13} according to which the effect of CDE on the degree of the supercluster straightness is stronger 
at earlier epochs (see Figure 8 in \citet{SL13}). Both of the CDE and the $f(R)$ gravity has the same effect of 
lowering the degree of the supercluster straightness but their evolution is directly the opposite, which implies 
that the degree of the supercluster straightness can be useful to distinguish between the two models.

\section{DISCUSSION AND CONCLUSION}\label{sec:conclusion}

A cosmological test of gravity has become a touchy topic. If an observable is to be regarded powerful in testing 
gravity, it should be sensitive enough not only to detect any little deviation of true gravity from GR but also to 
discriminate the effect of modified gravity from that of other energy contents such as coupled dark energy,  
warm dark matter, and etc. Since it was shown in plenty of literatures that the linear growth factor $D(z)$ 
and the Hubble expansion rate $H(z)$ in modified gravity models evolve differently from those in the standard 
$\Lambda$CDM or the dynamic dark energy models, much effort has been made to find 
observables which depend strongly on $D(z)$ and $H(z)$ 
\citep[e.g.,][and references therein]{linder05,zhang-etal07,HL07,wang08,zhao-etal09,SK09,shapiro-etal10}.

Very recently, however, \citet{wei-etal13} theoretically proved that it is practically impossible to distinguish among 
the scenarios of modified gravity, coupled dark energy and warm dark matter just by measuring  $D(z)$ and $H(z)$, 
because the predictions of the three scenarios for the evolution of those two quantities are effectively identical 
\citep[see also][]{wei-etal08}. They called this degeneracy among the three scenarios the "cosmological trinity". Given 
their claim,  an urgent work to undertake is to figure out which cosmological test has a power to break this inherent 
"trinity". 

Here, we have shown that the degree of the supercluster straightness has a capacity of completing  such a delicate 
mission.  We have found that the superclusters are significantly less straight in $f(R)$ gravity models than in the 
GR+$\Lambda$CDM. This effect is shown to become stronger in the models with larger values of 
$f_{R0}$. But, even the F6 model which is almost indistinguishable from the standard GR+$\Lambda$CDM model 
exhibits appreciable  difference in the degree of the supercluster straightness. The crucial implication of our 
result is that although the densities in the clusters are high enough to screen modified gravity inside the clusters, 
the intra-cluster force can still be unscreened. 

A comparison of our results with those obtained by \citet{SL13} that  the effect of cDE on the 
degree of the supercluster straightness becomes stronger at higher redshifts indicates that the degree of the 
supercluster straightness can be useful to discriminate the effect of modified gravity from that of cDE.
Although we have not investigated how the degree of the superclusters in the WDM models, it is very likely 
that the presence of WDM would make the superclusters more straight. As discussed in \citet{SL13}, the high 
peculiar velocities of dark matter particles plays a role in making the clustering of galaxy clusters more anisotropic.
That is, the WDM would make the superclusters more straight than the CDM. 

To use the degree of the supercluster straightness as a cosmological test of gravity, however, it will be much more 
desirable to have a theoretical framework within which the specific sizes of the supercluster spines can be 
evaluated for any cosmological models. Since we have obtained our results numerically from a N-body simulation 
which ran for a fixed model with specified values of the cosmological parameters, we do not know how much change it 
would cause to the degree of the supercluster straightness if different initial conditions were used as the simulation 
inputs. Given that the size of the main stem of a supercluster may correspond to the free streaming scale of the member 
clusters which can be treated as particles,  it might be possible to model how sensitive the free streaming scales of the 
clusters are to the initial conditions of the universe with the help of the Lagrangian perturbation theory. 

We would also like to mention that there is a good practical advantage of using the supercluster straightness as a probe 
of gravity. Unlike the other cosmological probes based on the galaxy clusters such as the cluster mass function, 
two-point correlations of the galaxy clusters and etc., it does not require accurate measurements of the masses of 
the galaxy clusters which are hard to achieve in practice. 
Once a sample of the galaxy clusters with masses larger than a certain threshold value is constructed,  
only information required to identify the superclusters and measure their straightness is the spatial positions of the 
sample clusters. Recently, a large sample of the galaxy clusters identified by the FoF algorithm from the Sloan Digital 
Sky Survey are available \citep[e.g.,][]{tempel-etal12}.  As mentioned in \citet{SL13}, once the redshift distortion 
effects are properly accounted for, our methodology can be readily applied to the observational datasets to 
determine the distribution of the supercluster straightness from the real universe. 
Our future work is in the direction of establishing a theoretical model for  the distribution of the supercluster straightness 
and applying our technique to real observations. 

\acknowledgments

We thank an anonymous referee for his/her helpful comments. 
JS and JL were financially supported by the Basic Science Research Program through the National 
Research Foundation of Korea(NRF) funded by the Ministry of Education (NO. 2013004372) and 
by the research grant from the National Research Foundation of Korea to the Center for 
Galaxy Evolution Research  (NO. 2010-0027910). BL thanks the support from the Royal Astronomical Society and 
Durham University. This work used the COSMA Data Centric system at Durham University, operated
by the Institute for Computational Cosmology on behalf of the STFC DiRAC HPC Facility (www.dirac.ac.uk). This 
equipment was funded by a BIS National E-infrastructure capital grant ST/K00042X/1, DiRAC Operations grant 
ST/K003267/1 and Durham University. DiRAC is part of the National E-Infrastructure.

\clearpage
\clearpage

\clearpage
\clearpage
\clearpage

\clearpage
\clearpage

\clearpage
\clearpage

\clearpage
\begin{deluxetable}{ccc}
\tablewidth{0pt}
\setlength{\tabcolsep}{5mm}
\tablecaption{Numbers of those supercluster spines with three or more nodes and their mean specific mass for the 
three models.}
\tablehead{model & $N_{\rm spine}$ & $\langle\tilde{M}_{\rm spine}\rangle$ \\
& & $[10^{13}\,h^{-1}M_{\odot}]$} 
\startdata
GR    &  $14204$  & $3.92$ \\
F6     & $15260$   & $3.80$ \\
F5    &  $16592$  & $4.18$ \\
\enddata
\label{tab:spine}
\end{deluxetable}

\end{document}